\documentclass[notnumberedtheorems,withtitlethanks]{actacybpress}

\usepackage{graphicx, algorithm, algorithmic, amsmath}

\usepackage{lineno}

\newcommand{\urlfootnote}[1]{\footnote{\url{#1}}}

\begin{document}



\title{Adding semantics to measurements:\\Ontology-guided, systematic performance analysis}
\headingtitle{Ontology-guided, systematic performance analysis}

\author{Attila Klenik\thanks{Hungary, Budapest, Budapest University of Technology and Economics, Department of Measurement and Information Systems, \email{\{attila.klenik, pataricza.andras\}@vik.bme.hu}, \orcid{https://orcid.org/\{0000-0003-2051-2823, 0000-0002-6516-129X\}}}, \and Andr\'{a}s Pataricza\thanksmark{1}}
\headingauthor{Attila Klenik, and Andr\'{a}s Pataricza}

\maketitle

\begin{abstract}
The design and operation of modern software systems exhibit a shift towards virtualization, containerization and service-based orchestration. Performance capacity engineering and resource utilization tuning become priority requirements in such environments.

Measurement-based performance evaluation is the cornerstone of capacity engineering and designing for performance. Moreover, the increasing complexity of systems necessitates rigorous performance analysis approaches. However, empirical performance analysis lacks sophisticated model-based support similar to the functional design of the system.

The paper proposes an ontology-based approach for facilitating and guiding the empirical evaluation throughout its various steps. Hyperledger Fabric (HLF), an open-source blockchain platform by the Linux Foundation, is modelled and evaluated as a pilot example of the approach, using the standard TPC-C performance benchmark workload.

\end{abstract}

\section{Introduction}
\label{sec:introduction}

The rapidly increasing number of IT service customers made the performance of such systems a high priority. Performant systems are not just a question of powerful hardware anymore, they also require the system-wide careful design of the software stack. The systematic detection and diagnosis of performance bottlenecks by analysing multi-dimensional measurement data becomes an integrated part of both the development and operational (DevOps) parts of the system life-cycle.

The industrialization of general-purpose data analysis resulted in typical standard workflows, like CRISP-DM~\cite{Wirth2000}, or ASUM-DM~\cite{Cerveira2018}. Such workflows are typically centered around the following high-level, domain-agnostic steps~\cite{Ardagna2017,Hashem2015,Streit2012}: \textit{data acquisition}; \textit{representation}; \textit{analysis}; \textit{visualization and reporting} -- with a proper model-driven engineering (MDE) support. Different performance analysis tasks -- such as bottleneck identification and latency anomaly root cause analysis -- can be considered domain-specific refinements~\cite{Kocsis2018} of the analysis step, defining further, embedded sub-workflows.

However, the \textit{technical metrology} of performance evaluation poses specific challenges. The technical systems under test (SUT) are usually very complex, both in the terms of their architecture and potential state space. Still, performance engineering became increasingly important, as many systems have to fulfill soft real-time requirements. Moreover, poor performance dimensioning (stemming from architectural design or misconfiguration) can lead to service-level violations, or the malfunctioning of the system, even in the case of short overloads.

\begin{figure}
\centering
\includegraphics[width=0.8\textwidth]{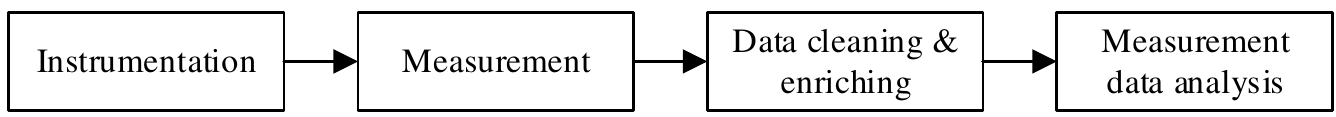}
\caption{Typical performance evaluation steps} \label{fig:performance-evaluation}
\end{figure}

The paper proposes an \textit{activity and observability-focused ontological approach} for the model-based guidance of SUT- and measurement-related, technical performance analysis tasks (Fig.~\ref{fig:performance-evaluation}): instrumentation; measurement; data cleaning and enriching; and measurement data analysis.

\textit{Instrumentation}, the insertion of sensors into the system, plays an important role in system \textit{observability}, i.e., the degree to which an observer can reconstruct the internal state of a system based on its outputs. However, sensor placement must balance multiple requirements: non-intrusivity whenever possible; development time/cost; and sufficient amount of resulting measurement data to work with. 

On one hand, increasing the number of sensors might provide a deeper insight into the system, but application-specific sensors require a careful development to assure the integrity of the measurement results without distorting temporal metrics. On the other hand, under-instrumentation confines the granularity of root cause analysis and consequently the indication for mitigating bottlenecks. Moreover, it can leave faulty behavior undetected. 

Correspondingly, instrumentation needs a careful trade-off between the relevance and redundancy of the measurements. The proposed approach aids the designer or analyst in formally arguing about the observability of the system, or in selecting a sufficient sensor placement.

During the \textit{measurement}, data acquisition has to cope with the heterogeneity of data sources generating observation logs in very different formats. Data source models~\cite{Klein2016,Streit2012} support the semantic fusion and representational homogenization of the sources and the following ETL (extract, transformation, load) steps. 

The proposed ontology guides the ETL process towards a \textit{relation-oriented and activity-focused representation} of measurement data, building on widely used concepts. The common format may serve as a gateway toward other temporal modeling frameworks (e.g. the OWL Time ontology\urlfootnote{https://www.w3.org/TR/owl-time/} from the World Wide Web Consortium), metrology-related technologies (e.g., the OpenTelemetry\urlfootnote{https://opentelemetry.io/} project from the Cloud Native Computing Foundation), or other analysis techniques (such as process mining~\cite{VanDerAalst2012}).

\textit{Clean and detailed data} is a prerequisite for many performance analysis tasks, such as bottleneck identification. Large-scale system observations constitute as \textit{big data}, but more importantly, as \textit{multi- or many-dimensional data}. Data is harvested from multiple layers of numerous system components, ranging from boundary-level response times to infrastructure-level resource utilization metrics. Bottleneck identification in such a context is a complex diagnostic process. It is a priori unknown how deep the analysis must drill down to uncover root causes of performance anomalies.

The proposed approach makes data validation a \textit{systematic process} by sharding and inspecting the data set based on the modeled activities and corresponding services. Thus data omission errors, for example, can be easily pinpointed even in larger data sets. Moreover, the activity and observability models coupled with various temporal rules provide a framework for automatically \textit{deriving further temporal information}, even if not explicitly observed.

The \textit{analysis} of the gathered multi-dimensional data necessitates \textit{exploratory data analysis} (EDA). EDA is, by its nature, a highly adaptive and iterative process for identifying a model of the system behavior. Usually, a domain expert is in charge of guiding the \textit{drill-down process} if something peculiar is detected from the point of view of the application. The exhaustiveness and quality of this exploration process heavily depend on the domain knowledge of the expert, the automation of the elementary steps, and a proper navigation along the process and the data~\cite{Streit2012,Garg2008,Perer2008,Yang2007}.

The \textit{hierarchical nature} of the proposed activity ontology and the corresponding service/deployment information make the drill-down process intuitive and systematic. The domain expert can dissect higher-level activities as needed, until a possible cause is found for a peculiar behavior. Then they can correlate the time range of the behaviour with various workload metadata and/or resource utilization metrics to find its root cause (may it be a bottleneck of the system, or a resource saturation issue). Furthermore, the drill-down steps are guided by concepts independent of the actual SUT, making the process reusable for different systems, a viable candidate for automation, or to be performed by a less experienced domain expert.

A complex case study demonstrates the benefits of the proposed approach:
\begin{itemize}
    \item The \textit{HLF blockchain} platform's consensus \textit{activities} and their \textit{observability} are modelled in a reusable and modular way.
    
    \item The activities of the standard TPC-C performance benchmark\urlfootnote{http://tpc.org/tpcc/default5.asp} are modeled and combined with the HLF model.
    
    \item The formal measurement inference capabilities of the ontology are demonstrated, coupled with a systematic data validation process.
    
    \item An ontology-guided, EDA-based, hierarchical bottleneck identification process is demonstrated on measurement data observed while executing the TPC-C workload (generated by the Hyperledger Caliper\urlfootnote{https://www.hyperledger.org/use/caliper} benchmarking tool) on HLF.
\end{itemize}


The paper is structured as follows. Sec.~\ref{sec:proposed-approach} introduces the proposed general approach for the performance analysis of complex systems. Sec.~\ref{sec:approaches} presents related MDE approaches for activity modeling and surveys the state of the art HLF performance researches. Sec.~\ref{sec:ontology-based-composite-activity-modeling} introduces the elements of the proposed ODK used for complex activity modeling and automatic observability inference, along with formal semantics. Sec.~\ref{sec:case-study-tpcc-benchmarking-hyperledger-fabric} presents the case study of compositional modeling of the HLF consensus process and the TPC-C benchmark execution using the ODK. Sec.~\ref{sec:systematic-analysis} demonstrates the various applicability of the resulting system models in aiding complex measurement data validation and analysis tasks. Sec.~\ref{sec:conclusion-and-future-work} concludes the paper.

\section{The proposed model-guided analysis approach}
\label{sec:proposed-approach}

The cornerstone of a performance analysis process is the observability of the activities performed by a system. On a high level, the beginning, duration, and end of system tasks are the basis of common metrics, like incoming task rate, throughput, and latency. Bottleneck identification, however, requires more interconnected data to work with, including the well-defined composition semantic of complex activities. Moreover, such observations are also crucial for building precise, well-parametrized models for efficient performance prediction~\cite{Becker2009,Garlan2004}.

Our contribution is an ontology-guided workflow for the \textit{systematic, drill-down performance analysis} of multi-dimensional measurement data. Moreover, the supporting \textit{ontology development kit} (ODK) is provided for ensuring the quality and sufficiency of measurement data, enhanced with composition semantics for facilitating bottleneck identification processes. The ODK supports the modeling of activities, their relations, and whether their execution is observable outside of the system. Furthermore, it provides a formal foundation for rigorous measurement data analysis task, e.g., bottleneck identification.

\begin{figure}
\centering
\includegraphics[width=0.8\textwidth]{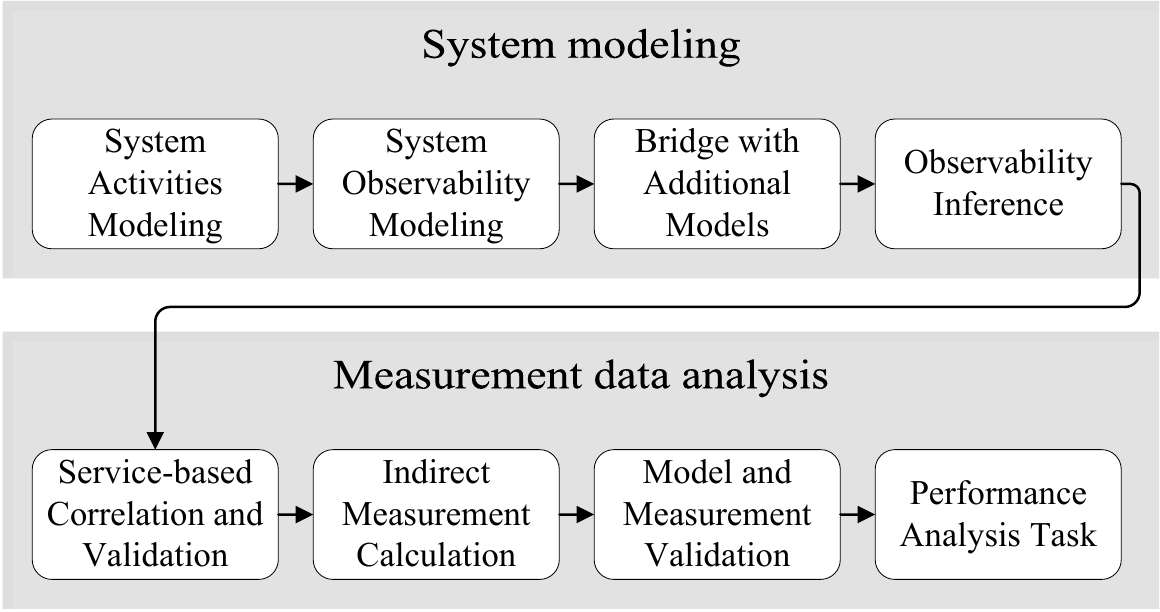}
\caption{The proposed workflow for model-guided performance evaluation} \label{fig:analysis-workflow}
\end{figure}

The proposed approach is outlined in Fig.~\ref{fig:analysis-workflow} and detailed in subsequent sections:
\begin{enumerate}
    \item Model the important activities of the system components, focusing on their relations and hierarchical composition.
    \item Model the explicit observability of activities to assist observability inference.
    \item Extend the model with additional elements (by bridging to other ontologies, for example) to support further design, DevOps, or analysis tasks, as needed.
    \item Automatically enrich the current model with additional observability information by using an OWL reasoner.
    \item Correlate and validate distributed request traces from the SUT to uncover data omission or similar errors.
    \item Calculate additional, indirect measurement using the ontology (or other derived) model as guide.
    \item Validate the conformance of the measurement data and the model to ensure the correctness of further analysis tasks.
    \item Perform the desired analysis task based on rigorously cleaned and validated data, and using the ontology model as guide.
\end{enumerate}

Following the outlined steps allows for a rigorous performance analysis of the SUT. Note that the model construction steps only need to be performed once (if done properly), then the component models can be reused and recombined to fit further performance analysis scenarios for different SUT setups. Moreover, the modeling and analysis parts of the workflow can be performed by different domain experts, lowering the entry barrier for the overall performance analysis of a given SUT. 
\section{Related work}
\label{sec:approaches}

The section presents the related work on creating activity execution models and surveys the state of the art regarding HLF performance evaluations. The limitations of the presented literature motivated our contribution to bring MDE approaches closer to the domain of performance evaluation of complex systems.

\subsection{Activity modelling}

System activities are usually observed through individual events (e.g., logs, notifications) or sensors. An important requirement of activity modeling -- relating to performance analysis -- is to allow the systematic reconstruction of detailed timelines from the available partial observations, facilitating data analysis. Furthermore, having well-defined modeling semantics and building blocks allow the assessment of a wide range of systems.

Our experience with EDA and bottleneck identification outlined the following requirements for model-based support:
\begin{itemize}
    \item \textit{formal modeling} of complex \textit{activity hierarchies} and relations;
    \item explicit modeling of system \textit{observability} (i.e., sensor placement);
    \item \textit{systematic derivation} of additional temporal knowledge;
    \item \textit{extensibility} for incorporating further service/infrastructure models;
    \item \textit{composability} and \textit{reusability} of different activity models.
\end{itemize}

Similar approaches exist in the domain of business process analysis using online analytical processing (OLAP) ~\cite{Abello2015,Neumayr2012,Niemi2010,Prat2012}. However, our approach has to comply with the additional requirements of technical metrology, like allowing the performance evaluation of general system activities despite limited observability of tasks, and facilitation of the adaptation of metrology principles.

Modeling the execution of activities also has a long tradition in software development, both as design phase artifacts for validation, and as inputs to automatic task orchestration systems. Business process models (building on the BPMN\footnote{https://www.omg.org/spec/BPMN/2.0/} standard) or activity diagrams in UML\footnote{https://www.omg.org/spec/UML/} or SysML\footnote{https://www.omg.org/spec/SysML/} are prime examples of high-level activity modeling languages. 

Such visual languages facilitate the modeling of activity control flows, imposing certain temporal constraints (e.g., activity $A$ must be executed \textit{before} activity $B$). However, the enforcement of such constraints must be validated during analysis time or runtime. Such validation necessitates the detailed observation of activities to allow rigorous temporal constraint checks. Moreover, the available high-level languages lack an intuitive support of modeling observability.

Ontology-like formal approaches also gain ground in general system modeling tasks (e.g., the upcoming OMG SysML v2 Kernel Modeling Language\footnote{https://github.com/Systems-Modeling/SysML-v2-Release}), thus our contribution relies on ontologies, preparing for future interoperability. Knowledge representation-based approaches can also aid the visual analysis of network traffic~\cite{Xiao2006} or the semantic fusion of data originating from different sources~\cite{Wun2007}. Moreover, ontology-based approaches can reason about the occurrence of composite activities~\cite{Okeyo2012,Chen2012,Helaoui2011,Meditskos2013,Riboni2011}.

The referenced activity modeling works have several elements in common. They utilize Allen’s interval algebra~\cite{Allen1983} for describing temporal relations, allowing bridging to other similar solutions. However, they reverse-engineer/infer the activity model based on the observation of performed activities, similarly to process mining~\cite{VanDerAalst2012}. Model mining is unavoidable in contexts where the ''schedule'' of executed activities is non-deterministic, such as in smart homes or in smart warehouses~\cite{Chen2012,Riboni2011}.

However, when the execution of activities must conform to a predefined specification, model mining becomes unnecessary. The paper proposes a \textit{model-first} approach to construct an ontology-based composite activity model, which will later provide a strong foundation for the systematic performance evaluation and bottleneck analysis of the target system. Accordingly, the model becomes an input to the analysis tasks, and not an output.

\subsection{Hyperledger Fabric performance analysis}

The complex consensus process of HLF~\cite{Androulaki2018} (detailed and modeled in Sec.~\ref{sec:case-study-tpcc-benchmarking-hyperledger-fabric}) made its performance evaluation a hot research topic. Related works can be divided mainly into the following categories based on their goals:
\begin{enumerate}
    \item\label{fab-rel-cat-1} Performance \textit{evaluation and characterization}:  \cite{Pongnumkul2017,Baliga2018,Thakkar2018,Nasir2018,Gupta2018,Takeshi2018,Sharma2018,Hao2018,Wang2019,Kuzlu2019,Nguyen2019,Inagaki2019,Nguyen2019b,Androulaki2019,Foschini2020,Wang2020,Shalaby2020,Bergman2020}
    
    \item\label{fab-rel-cat-2} Performance \textit{optimization}: \cite{Thakkar2018,Gorenflo2019,Javaid2019,Nakaike2020}
    
    \item\label{fab-rel-cat-3} Formal consensus \textit{modelling}: \cite{Sukhwani2017,Sukhwani2018,Jiang2020,Yuan2020,Xu2021}
\end{enumerate}

Category~\ref{fab-rel-cat-1} receives most of the attention, which is identifying the performance characteristics of HLF. The evaluations employ empirical sensitivity analyses to measure the change in key performance indicators (such as throughput and end-to-end latency) when applying different network scales, configurations, and workloads.

The concern of Category~\ref{fab-rel-cat-2} is the performance enhancement of HLF. Researches either transparently optimize certain consensus steps or propose changes to the architecture (and correspondingly the consensus process) itself. The researches of Category~\ref{fab-rel-cat-2} also rely on empirical performance analysis to confirm bottlenecks and evaluate the effectiveness of optimizations.

Works in Category~\ref{fab-rel-cat-3} build formal behavior models of the consensus process. Model parameter identifications also rely on empirical performance evaluations. Finally, the parameterized model allows for cost-efficient sensitivity analyses capable of covering a large configuration and parameter space, without actual further empirical analyses.

A common requirement for all three categories is the rigorous empirical performance evaluation of HLF based on the analysis of measurement data. Superficial analyses may lead to incorrect hypotheses or misidentified model parameters, invalidating the results of the evaluation. 

Accordingly, a \textit{systematic}, \textit{rigorous}, and \textit{easy to follow} analysis process (even for complex systems) is needed to achieve relevant results. Moreover, the \textit{correctness} and \textit{richness} of measurement data can further increase the quality of gained insights.

\section{Activity and observability modeling framework}
\label{sec:ontology-based-composite-activity-modeling}

The section introduces the formal foundations and building blocks of the proposed ODK for constructing complex activity models. Moreover, it details the observability modeling and automatic observability inference mechanisms that are the cornerstones of a rigorous performance data analysis.

\subsection{Formal foundations}
\label{subsec:formal-foundations}

The ODK is constructed using the Web Ontology Language\footnote{https://www.w3.org/TR/owl2-syntax/} (OWL2), adhering to some constraints (resulting in an OWL-DL ontology) that make the OWL direct semantics compatible with the model-theoretic semantics of the $\mathcal{SROIQ}$ description logic~\cite{Horrocks2006}. This restriction provides useful computational properties for the language, backed by extensive literature and tooling support, such as OWL-DL reasoners~\cite{Parsia2017}.

OWL2 provides facilities such as object and data properties, literals, individuals, and classes to model relations among different concepts and resources. Classes can have associated relationship constraints that must hold for every individual belonging to the class. It is important to note that OWL employs the open-world assumption, meaning that if something is not asserted as knowledge, it is taken as unknown, rather than as untrue. The OWL2 structural specification\footnote{https://www.w3.org/TR/owl2-syntax/} further details the available language constructs and their meaning.

The paper utilizes the OWL-DL notations of Table~\ref{table:owl-dl-notations} to describe the elements of the ODK and their semantics.

\begin{table}
    \caption{OWL-DL notations}\label{table:owl-dl-notations}
    \centering
    \begin{tabular}{|l|l||l|l|}
        \hline
        OWL construct & Notation & OWL construct & Notation\\
        \hline
        \hline
        $Class$ & $C1, C2$ & $SubClassOf$ & $C1 \subseteq C2$\\
        $IntersectionOf$ & $C1 \cap C2$ & $UnionOf$ & $C1 \cup C2$\\
        $Thing$ & $T$ & $Property$ & $P$\\
        $PropertyRange$ & $T \subseteq \forall P.C1$ & $EquivalentClass$ & $C1 \equiv C2$\\
        $AllValuesFrom$ & $C1 \subseteq \forall P.C2$ & $SomeValuesFrom$ & $C1 \subseteq \exists P.C2$\\
        \hline
    \end{tabular}
\end{table}

The temporal constructs of the ODK build on Allen’s interval algebra. Let us consider activity instances $a = (a^b, a^e, a^d) \in A$ of an activity class with a beginning time instant $a^b \in \mathbb{N}^+$, an ending time instant $a^e \in \mathbb{N}^+$, and a non-zero duration $a^d \in \mathbb{N}^+$, where $a^b < a^e$, and $a^b + a^d = a^e$ for every $a \in A$, measured on a logical clock for the simplicity of the notation. 

If a property (i.e., directed relation) $P$ holds between activity instances $a$ and $b$, we denote it by $a \in P.b$, where $a \in A$, $b \in B$ activity classes. The shorthand notation $A \subseteq P.B$ specifies the relation $P$ as constraint between activity classes $A$ and $B$, implying $\forall a \in A: \exists b \in B, a \in P.b$. 

The ODK defines the following Allen interval relations as OWL properties: $after$, $before$, $meets$, $metBy$, $starts$, $startedBy$, $finishes$, and $finishedBy$. Accordingly, if an activity type $A$ is always followed by an activity type $B$, the axiom $A \subseteq meets.B$ will be part of the ontology. Note, that the ODK contains only the Allen relations that provide precise or useful activity composition semantics. Accordingly, the $during$, $overlaps$, and $equal$ relations (and their inverses) are not utilized directly, but can be derived from the modeled relationships in a straightforward way.

\subsection{Component overview}
\label{subsec:component-overview}

\begin{figure}
\centering
\includegraphics[width=\textwidth]{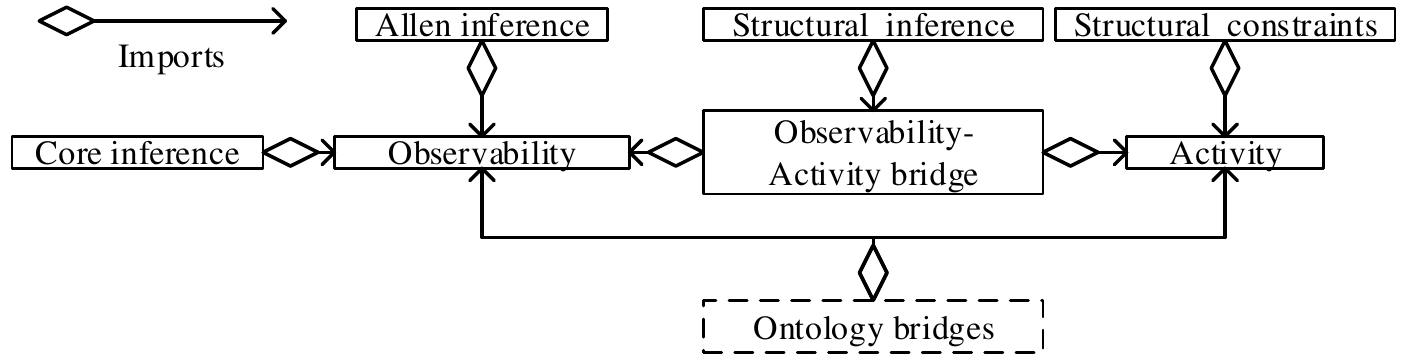}
\caption{Ontologies in the ODK} \label{fig:odk-components}
\end{figure}

The ODK contains a hierarchy of smaller ontologies -- each with well-defined responsibilities -- to promote composability (Fig.~\ref{fig:odk-components}).

\begin{figure}
\centering
\includegraphics[width=0.6\textwidth]{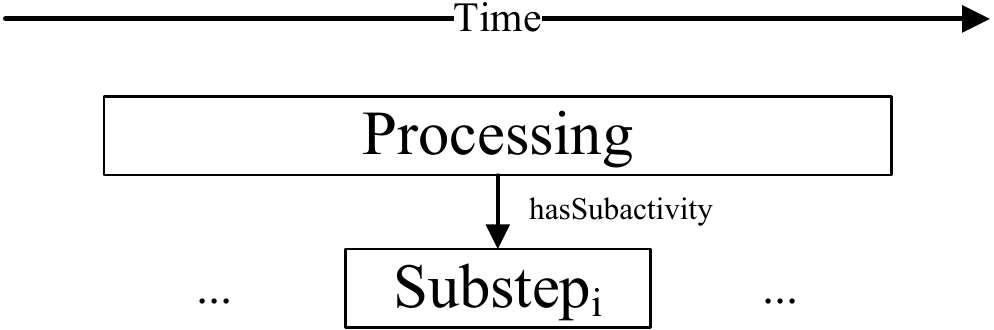}
\caption{Example of a partial activity decomposition} \label{fig:activity-example-1}
\end{figure}

The $Activity$ ontology (Sec.~\ref{subsec:activity-composition}) allows the modeling of system activity relations. For example, the following set of assertions partially describe an activity decomposition (Fig.~\ref{fig:activity-example-1}): $Processing \subseteq SequentialActivity$, $Substep_i \subseteq AtomicActivity$, and $Processing \subseteq hasSubactivity.Substep_i$. 

The $Observability$ ontology (Sec.~\ref{subsec:modeling-observability}) provides classes to ''annotate'' the activities with further information regarding their degree of observability. For example, $Processing \subseteq EndMeasured$ denotes that the end of $Processing$ activities are explicitly observed/measured through logs, or system events.

The $Structural\ constraints$ ontology provides well-formedness axioms for activity composition. The open-world semantic of OWL2 makes it cumbersome to convey traditional (closed-world) modelling intentions to a set of ontology axioms. For example, it is not enough to just state that a subactivity is the starting activity of its parent. Correct modeling also requires the statement that the starting activity is not preceded by any other activity (otherwise it could not be the first subactivity of its parent). 

The structural constraints ontology provides several axioms that can automatically detect (using an ontology reasoner) such inconsistencies or potentially missing axioms. The description of constraints, however, is outside the scope of this paper.

The $inference$ ontologies (Sec.~\ref{subsec:observability-inference}) extend the observability ontology with equivalence axioms that can automatically flag (during reasoning) activity classes based on their degree of observability. Such observability flags are propagated during reasoning along the activity relations, resulting in a complete observability description of the entire activity hierarchy.

The structural constraint and inference ontologies can be referred to as \textit{aspect ontologies} in general: they separate orthogonal modelling concerns in a modular way and can be used to enrich a base model (similarly to aspects is aspect-oriented programming). Accordingly, a modeler can work using light-weigh and simple ontology concepts (activities and observability), and only perform possibly heavyweight computations/reasoning periodically by including the aspects. 

\begin{figure}
\centering
\includegraphics[width=\textwidth]{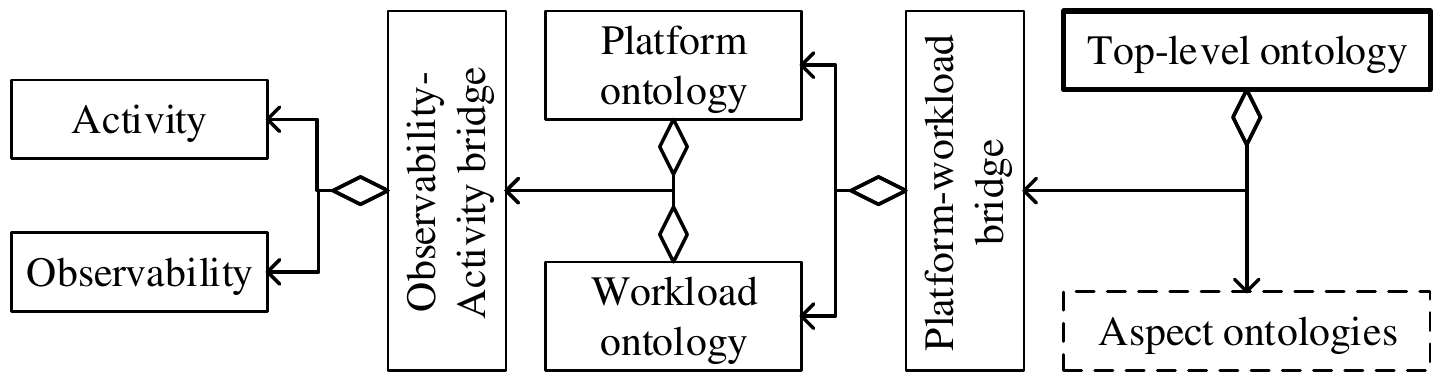}
\caption{Ontology structure of a multi-component use case} \label{fig:general-use-case-structure}
\end{figure}

The recommended modelling approach of the ODK is to shard the complete system model into smaller ontologies (Fig.~\ref{fig:general-use-case-structure}) for maximum flexibility and reusability.

A platform ontology models the composition of executed activity steps and their observability utilizing the ODK core vocabulary. A specific platform is usually just a means to execute a higher-level business scenario, which steps should be modeled as a platform-independent workload ontology whenever possible. This separation allows the flexible evaluation of different architectural/platform design choices by providing a platform-workload bridging ontology for specific scenarios. 

The final element of the stack is a top-level (possibly automatically constructed) ontology that unites the pure system model with the chosen aspects of the ODK. Various OWL-DL reasoners can validate and enrich the top-level ontology, resulting in a detailed knowledge representation of the system that will serve as a basis for later performance analysis tasks. A concrete case study following the presented approach is detailed in Sec.~\ref{sec:case-study-tpcc-benchmarking-hyperledger-fabric} through modeling TPC-C benchmark execution on HLF networks.

\subsection{Modeling activity hierarchies}
\label{subsec:activity-composition}

\begin{figure}
\centering
\includegraphics[width=\textwidth]{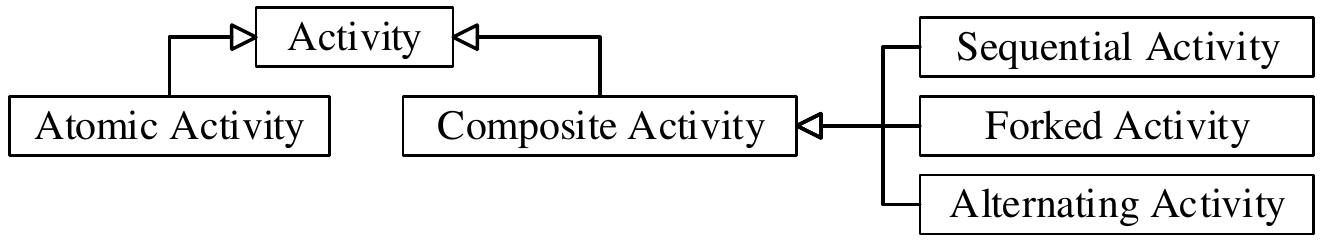}
\caption{Hierarchy of activity types} \label{fig:activity-hierarchy}
\end{figure}

$Activity$ ($ACT$) hierarchies are defined with atomic (''leaf'') elements and composite elements supporting further refinement (Fig.~\ref{fig:activity-hierarchy}). The ODK provides the following $Activity$ subclasses for modeling activity composition through subsumption relations:
\begin{itemize}
    \item $AtomicActivity$ ($AA$) represents elementary steps without further refinement;
    \item $CompositeActivity$ ($CA$) allows further refinement of activities through the following subclasses, representing different composition semantics:
    \begin{itemize}
        \item $SequentialActivity$ ($SA$) allows refinement into a sequence of subactivities;
        \item $ForkedActivity$ ($FA$) allows refinement into parallel subactivities;
        \item $AlternatingActivity$ ($TA$) allows refinement into subactivities without additional control flow constraints.
    \end{itemize}
\end{itemize}

The core classes on the same hierarchy level are disjoint ($AA \cap CA = \emptyset$, $SA \cap FA \cap TA = \emptyset$). However, uncategorized activities and additional composition semantics are allowed to promote extendability, i.e., $AA \cup CA \neq ACT$ and $SA \cup FA \cup TA \neq CA$.

The following high-level relations (OWL object properties) provide the basis for constructing complex hierarchies of activities:
\begin{itemize}
    \item $Substep_i \subseteq hasParentActivity.Parent$, denoting that activity type $Substep_i$ has a parent (encapsulating) activity of type $Parent$.
    \item $Parent \subseteq hasSubactivity.Substep_i$, denoting that activity type $Parent$ has a subactivity (a refined substep) of type $Substep_i$.
    \item $Substep_i \subseteq hasSiblingActivity.Substep_j$, denoting that $Substep_i$ has the same parent activity type as $Substep_j$, i.e., $\exists Parent$ such that $Substep_i \subseteq hasParentActivity.Parent$, and $Substep_j \subseteq hasParentActivity.Parent$
\end{itemize}

The composite activity subclasses denote the typical (de)composition constructs for activity executions:

\begin{description}
\item[Sequential activities] ($SA$) group together a sequence of subactivities that follow traditional sequential execution semantics. Moreover, refined relations are introduced (with Allen interval-like semantics, as mapped in Table~\ref{table:sequential-refinements}) to further enrich parent-subactivity and sibling relations. The ''synonyms'' for the Allen relations were introduced to hint at the compositional nature of the activities (and not just their relative temporal placement), aiding modelers with traditional activity modeling backgrounds. 

Note, that ''gapped'' relations indicate an incomplete timeline, hindering later analyses, and probably warranting additional instrumentation. However, the ODK inference rules can be easily extended to detect ''unknown'', albeit observable activities whenever possible.

\begin{table}
    \caption{ODK and Allen relation mappings for sequential composition}\label{table:sequential-refinements}
    \centering
    \begin{tabular}{|c|c||c|}
        \hline
        Parent Relation $\supseteq$ & Subrelation $\equiv$ &  Allen relation\\
        \hline
        \hline
        $hasParentActivity$ & $startsParentActivity$ & $starts$ \\ 
        & $finishesParentActivity$ & $finishes$ \\
        $hasSubactivity$ & $startedBySubactivity$ & $startedBy$ \\
        & $finishedBySubactivity$ & $finishedBy$ \\
        $hasSiblingActivity$ & $hasImmediatePredecessorActivity$ & $metBy$ \\ 
        & $hasImmediateSuccessorActivity$ & $meets$\\
        & $hasGappedPredecessorActivity$ & $after$ \\
        & $hasGappedSuccessorActivity$ & $before$\\
        \hline
    \end{tabular}
\end{table}

\item[Forked activities] ($FA$) group together parallel subactivities that are executed independently of each other. An associated synchronization/join semantic class ($\subseteq hasSyncSemantic.SyncSemantic$) can be used to model the condition when the parent activity is deemed finished.

The ODK defines two synchronization semantics: when \emph{all} ($WaitForAll \subseteq SyncSemantic$), or \emph{any} ($WaitForAny \subseteq SyncSemantic$) of the subactivities must finish to consider the parent activity done. Extending ontologies can define further semantics, e.g., waiting for the majority of subactivities.

\item[Alternating activities] ($TA$) are decomposed into a set of subactivities, disregarding control flow restrictions in cases when the control flow of subactivities is irrelevant. $TA$ is a tool of abstraction for concentrating only on the ''weight'' (i.e., duration) of a subactivity, and not on its scheduling. 

A typical use case is the modeling of in-process execution times and database access times of a task, disregarding execution semantics among the substeps: $Task \subseteq TA$, $InProcExec, DbAccess \subseteq hasParentActivity.Task$. Modeling the exact activity flow of computation and database access can be cumbersome for some use cases. Moreover, it may be sufficient during performance analysis to consider only the time/duration spent with each processing types, instead of  focusing on their exact, possibly rapidly alternating order. 
\end{description}
\subsection{Modeling observability}
\label{subsec:modeling-observability}

Once the activity model is complete, the next step is modeling which activity temporal aspects (beginning, duration, and/or end) are measured directly in the system (i.e., modeling the placement of sensors and instrumentation) using the $Observability$ ontology concepts.

\begin{figure}
\centering
\includegraphics[width=\textwidth]{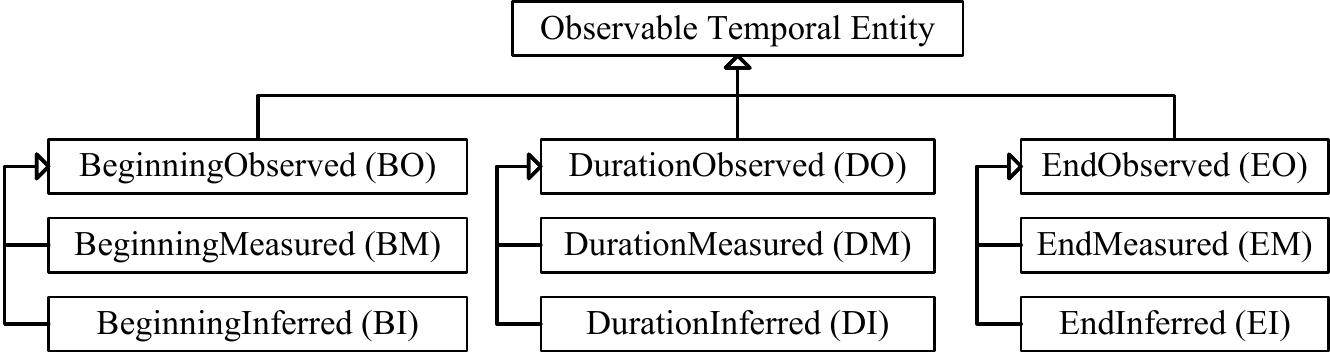}
\caption{Observability ontology components, with abbreviations} \label{fig:observability-cropped}
\end{figure}

The core concepts can be grouped into three main categories (Fig.~\ref{fig:observability-cropped}): 
\begin{enumerate}
    \item observable data in the abstract sense (e.g., $Task \subseteq EndObserved$), denoting that the temporal data is available in some way (measured or inferred);
    
    \item directly measured data (e.g., $Task \subseteq EndMeasured \subseteq EndObserved$), denoting that the data is explicitly measured;
    
    \item and inferred data (e.g., $Task \subseteq Rule \subseteq EndInferred \subseteq EndObserved$).
\end{enumerate}
 
 The following class abbreviations are used in some places for readability (Fig.~\ref{fig:observability-cropped}): $BO$, $BM$, $BI$, $DO$, $DM$, $DI$, $EO$, $EM$, and $EI$.

The modeler must ''annotate'' each activity class if one or more of its temporal aspects are directly measured in the system. For example, if the system logs the end time of an activity $Processing$, then the modeler can add the following axiom to the ontology: $Processing \subseteq EndMeasured$, also implicitely stating that $Processing \subseteq EndMeasured \subseteq EndObserved$. Such annotations will serve as a priori knowledge to the reasoner later. Moreover, additional instrumentation knowledge can be encoded in the ontology if modelers subsume the $*Measured$ classes (e.g., name of the logging component, format, reference to source code, etc.).

The general classes for observable data ($BO$, $DO$, and $EO$) provide an abstraction layer that hides the exact source of observability. Inference rules will reference this abstract level to handle and propagate explicit and inferred observability uniformly (Sec.~\ref{subsec:observability-inference}).

The $BI$, $DI$, and $EI$ classes are the extension points of the observability ontology, i.e., the superclasses for implementing observability inference rules, as detailed next.

\subsection{Observability inference}
\label{subsec:observability-inference}

Given a partially observable activity model, an OWL-DL reasoner can infer further observable temporal aspects utilizing inference rules based on Allen interval and structural relations.

The observability inference is implemented using the class equivalence construct of OWL2. The rules are modeled as OWL classes (e.g., $RuleX \subseteq EndInferred$) with corresponding equivalence axioms as \emph{criteria} (describing an anonymous class in OWL in the form of $RuleX \equiv criteria$). The axioms of criteria usually encode some kind of temporal data propagation rule among activities, while referencing the abstract observability of the involved activities.

When a reasoner infers that an activity type $Processing$ satisfies the criteria (i.e., subsumes the corresponding anonymous class), $Processing$ becomes part of the class hierarchy of the corresponding $*Observed$ class. E.g., if we have a rule ($RuleX$) about inferring the end time of an activity based on some $criteria$ (i.e., $RuleX \equiv criteria$), then the following axiom will be added the ontology if $Processing$ ''matches'' $criteria$: 

$$Processing \subseteq criteria \equiv RuleX \subseteq EndInferred \subseteq EndObserved$$

Accordingly, $Processing$ will be categorized as an $EndObserved$ class, allowing the propagation of the newly inferred knowledge through other rules, continuing until no new knowledge can be inferred.

\subsection{Inference rules}

The inference mechanism is demonstrated through the simple constraint between the beginning time, duration, and end time of any activity instance: $a^b + a^d = a^e, \forall a \in A \subseteq ACT$. This constraint is the basis of the three core inference rules (Eqs.~\ref{eq:inference-1}--\ref{eq:inference-3}) provided by the ODK: if two of the temporal aspects are observable, then the third is inferrable. Rules are encoded through equivalent class axioms and subsume the proper $BI$, $DI$ or $EI$ inference extension points.

\begin{equation}
\begin{aligned}
    A \subseteq (DO \cap EO) \equiv Rule1 \Longrightarrow A \subseteq Rule1 \subseteq BI \subseteq BO\\
\end{aligned}
\label{eq:inference-1}
\end{equation}

\begin{equation}
\begin{aligned}
    A \subseteq (BO \cap EO) \equiv Rule2 \Longrightarrow A \subseteq Rule2 \subseteq DI \subseteq DO\\
\end{aligned}
\label{eq:inference-2}
\end{equation}

\begin{equation}
\begin{aligned}
    A \subseteq (BO \cap DO) \equiv Rule3 \Longrightarrow A \subseteq Rule3 \subseteq EI \subseteq EO 
\end{aligned}
\label{eq:inference-3}
\end{equation}

The ODK contains numerous additional inference rules based on Allen interval and structural relations. The following conjunctions of criteria (in the form of $A \subseteq (criteria_1 \cup ... \cup criteria_n) \Longrightarrow A \subseteq BO/DO/EO$) succinctly encode the additional rules for inferring beginnings, durations, and ends, respectively:

\begin{equation}
\begin{aligned}
    A \subseteq ((FA \cap \forall hasSubactivity.BO) \cup (\exists starts.BO) \cup (\exists startedBy.BO) \cup\\
    (\exists metBy.EO) \cup (\exists hasParentActivity.(BO \cap FA)))
    \Longrightarrow  A \subseteq BO
\end{aligned}
\end{equation}

\begin{equation}
\begin{aligned}
    A \subseteq ((\exists hasParentActivity.(TA \cap DO) \cap \forall hasSiblingActivity.DO) \cup\\
    (TA \cap \forall hasSubactivity.DO)) \Longrightarrow A \subseteq DO
\end{aligned}
\end{equation}

\begin{equation}
\begin{aligned}
    A \subseteq ((\exists meets.BO) \cup (\exists finishes.EO) \cup (\exists finishedBy.EO) \cup\\
    (FA \cap \forall hasSubactivity.EO) \Longrightarrow A \subseteq EO
\end{aligned}
\end{equation}

The ODK allows the declaration of additional rule classes by simply subsuming the appropriate $*Inferred$ classes.

Note that an activity $A$ can match multiple rules. For example, if two temporal aspects are observable, $A$ matches one of Eqs.~\ref{eq:inference-1}--\ref{eq:inference-3}. However, now all three aspects are observable, so $A$ matches all three rules. In general, the matching inference rules between temporal data define a data flow network, facilitating various data analysis tasks, as detailed in Sec.~\ref{sec:systematic-analysis}.

\subsection{ODK extendability}
\begin{figure}
\centering
\includegraphics[width=\textwidth]{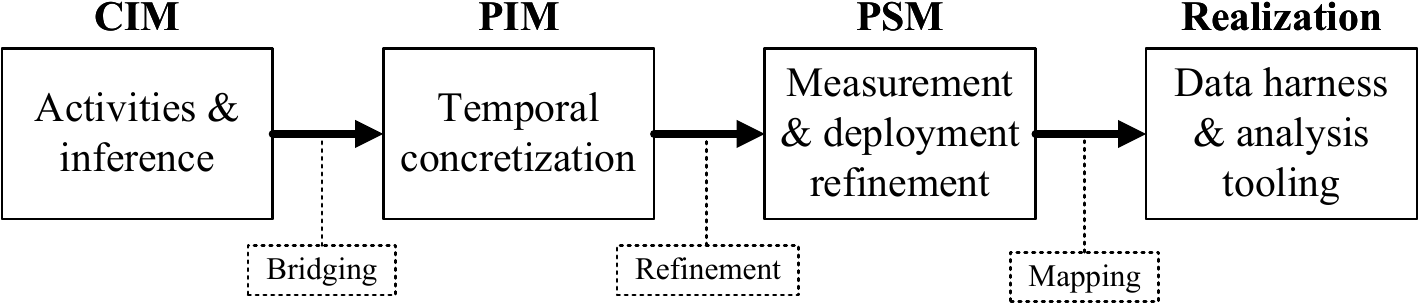}
\caption{Envisioned MDE flow of activity modelling}
\label{fig:model-flow}
\end{figure}

The ODK operates with high-level and abstract concepts in order to allow extendability with additional concepts, increasing the flexibility and usability of the model in subsequent analysis tasks. Fig.~\ref{fig:model-flow} contextualizes the different ODK capabilities in the typical MDE worklfow.

The core activity concepts and inference rules comprise a computation-independent model (CIM) for describing observability in a temporal representation-agnostic way. The models at this level (e.g., in Sec.~\ref{sec:case-study-tpcc-benchmarking-hyperledger-fabric}) only state knowledge like the beginning of a \textit{Processing} activity is measured. Information about \textit{how} that measurement is acquired, and in \textit{what format}, is omitted. Moreover, inference rules define only the data dependency of calculated measurements -- again, omitting the exact computational steps. 

The first step towards an actual realization of the analysis process is to enrich the core model with temporal data and corresponding relations. E.g., the OWL Time ontology\urlfootnote{https://www.w3.org/TR/owl-time/} defines \textit{Interval}s as ''a temporal entity with an extent or duration.'' Furthermore, the ontology defines the \textit{hasBeginning} relation (among others) between temporal entities (such as intervals) and arbitrary time instants. A simple \textit{bridging} between the two ontologies (e.g., $Activity \equiv Interval$) enriches the activities with actual temporal data formats. The associated temporal data concretize the manner of measurement calculations, but still neglects the exact source and harness of measurement data, thus acting as a platform-independent model (PIM).

Two ODK aspects support the refinement of PIMs into platform-specific models (PSM). On one hand, additional ontologies can refine classes like \textit{BeginningMeasured} to inlcude the source of the measurement data. For example, an extending ontology could define a \textit{BeginningLogged} subclass of \textit{BeginningMeasured}, providing details about the log format, source software component, and the semantic structure of the message, all aiding the extraction of measurement information in a log processing pipeline. 

On the other hand, the ODK provides an \textit{executedBy} relation to associate an \textit{Activity} type with a \textit{Service} type (e.g., \textit{Endorsement} activities are \textit{executedBy} \textit{PeerService}s). Extending ontologies can build on this relation to further model the deployment information related to a certain environment where the SUT is operated. For example, a deployment ontology could maintain information about a HLF network, where each service instance is located on a certain Kubernetes\urlfootnote{https://kubernetes.io/} node, in a cluster comprised of several virtual machines, hosted on specific hardware components.

Finally, a technology stack realizing the actual data analysis flow can utilize all levels of the final, rich model to uncover the root cause of an anomalous activity duration/latency (partially demonstrated in Sec.~\ref{subsec:bottleneck-id}), even if stemming from the lowest level of the infrastructure.
\section{Case study: Modeling TPC-C on Fabric}
\label{sec:case-study-tpcc-benchmarking-hyperledger-fabric}

Performance benchmarks serve as platform-agnostic workload specifications representative for a given domain, facilitating the comparison of different backend platform implementations under reproducible conditions. The benchmark plays the role of a platform-independent model (PIM) in MDE terminology, while the emulated clients and database engine make it platform-specific. The section introduces a compositional model of the TPC-C workload executed on HLF, using the presented ODK concepts as case study.

\subsection{Modeling the TPC-C benchmark}

TPC-C is a mature online transaction processing (OLTP) benchmark inspired by the typical activities of a wholesale supplier. TPC-C uses a mix of five transaction types -- with varying complexity -- to be executed against a rich database schema (HLF in the case study).

The execution of a TPC-C transaction by an emulated client has the following, strictly sequential composition of steps: (1) the client selects a transaction type ($Menu\ selection$); (2) then fills the required inputs for the request ($Fill\ inputs$); (3) then the database engine executes the transaction ($Execute\ TX$); (4) and finally the client takes some time to think about the next transaction ($Think\ time$) before starting the next cycle. The model of this client cycle plays the role of the workload ontology in Fig.~\ref{fig:general-use-case-structure}.

\begin{figure}
\centering
\includegraphics[width=\textwidth]{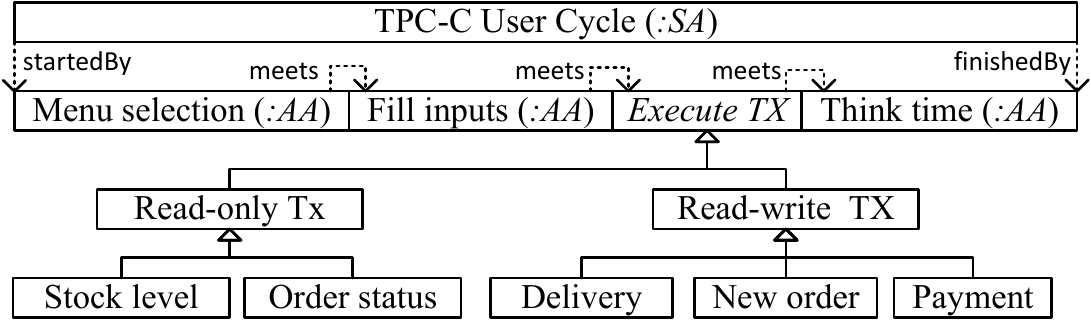}
\caption{TPC-C transaction execution scheduling} \label{fig:tpcc-high-level}
\end{figure}

Accordingly, the activity model (Fig.~\ref{fig:tpcc-high-level}) declares a top-level/root sequential activity, having four subactivities. The $Menu\ selection$, $Fill\ inputs$, and $Think\ time$ subactivities are atomic activities that simply emulate user behavior through artificial delays with specified distributions.

The exact composition of the $Execute\ TX$ activity is specific to the database engine, thus its type does not subsume any of the $Activity$ ontology classes. The exact type binding is the task of a platform-workload bridging ontology that maps the platform request execution activities to the $Execute\ TX$ activity.

The TPC-C transaction types are further categorized based on whether they are read-only, or read-write requests, making the bridging easier to platforms that differentiate between the execution of the two categories (like HLF does). 

\subsection{Modeling the Hyperledger Fabric consensus}

During the benchmark measurement, a HLF network served as the ''database engine.'' The novelty of HLF is its execute-order-validate style consensus mechanism, breaking with the traditional order first approaches~\cite{Androulaki2018}. However, its performance characterization is still incomplete. The case study models the detailed HLF consensus mechanism, enriched with client-side observations provided by the Hyperledger Caliper workload generator.

\begin{figure}
\centering
\includegraphics[width=\textwidth]{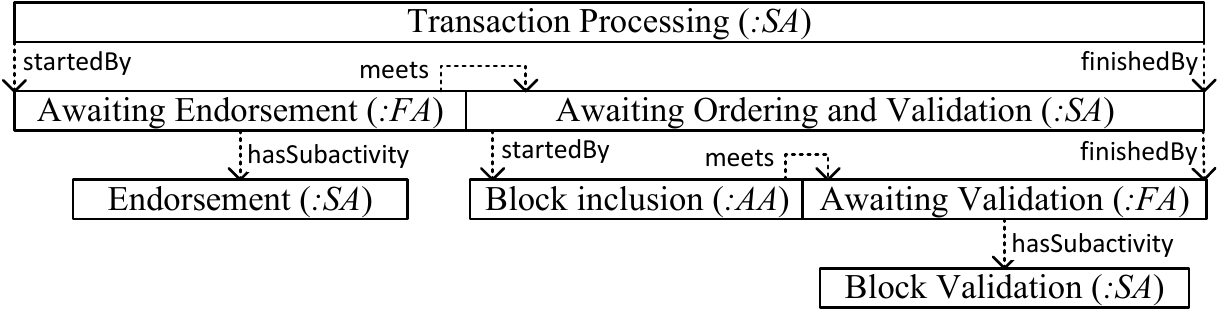}
\caption{High-level steps of the HLF consensus} \label{fig:hlf-high-level}
\end{figure}

The concepts and consensus steps of HLF are detailed in \cite{Androulaki2018} or in the official documentation.\urlfootnote{https://hyperledger-fabric.readthedocs.io/en/release-1.4/txflow.html} The section focuses only on the composition of activities (and not on their technical descriptions) to demonstrate that deep domain knowledge is not required during the guided performance analysis tasks. Note that creating the model, however, requires familiarity with the modeled platform, but ideally it is the responsibility of the designers or platform experts to create such a model. 

Fig.~\ref{fig:hlf-high-level} details the high-level, sequential steps of the HLF transaction life-cycle. The model plays the role of the platform ontology in Fig.~\ref{fig:general-use-case-structure}.

Clients first assemble and send a transaction proposal to one or multiple peers for \textit{parallel} simulation/endorsement and wait for \textit{all} results (\textit{Awaiting Endorsement} activity) to arrive, modeled by an associated \textit{WaitForAll} synchronization semantic. Once the results are available, the client then sends them to the ordering service and waits for a notification from the network that the transaction was successfully committed or not (\textit{Awaiting Ordering and Validation} activity).

The ordering and validation phase is modelled by two consecutive subactivities: \textit{Block inclusion} and the client \textit{Awaiting Validation} from \textit{any} peer (denoted by a \textit{WaitForAny} synchronization semantic). It is important to note, that the \textit{Awaiting Validation} activity is not a dedicated, explicitly observable activity of the client. It is artificially introduced for convenience to separate the ordering and the validation steps for detailed analysis. This choice demonstrates that the activity model is constructed in a way to facilitate detailed performance analysis, rather than be a technically faithful representation of the platform.

\begin{figure}
\centering
\includegraphics[width=\textwidth]{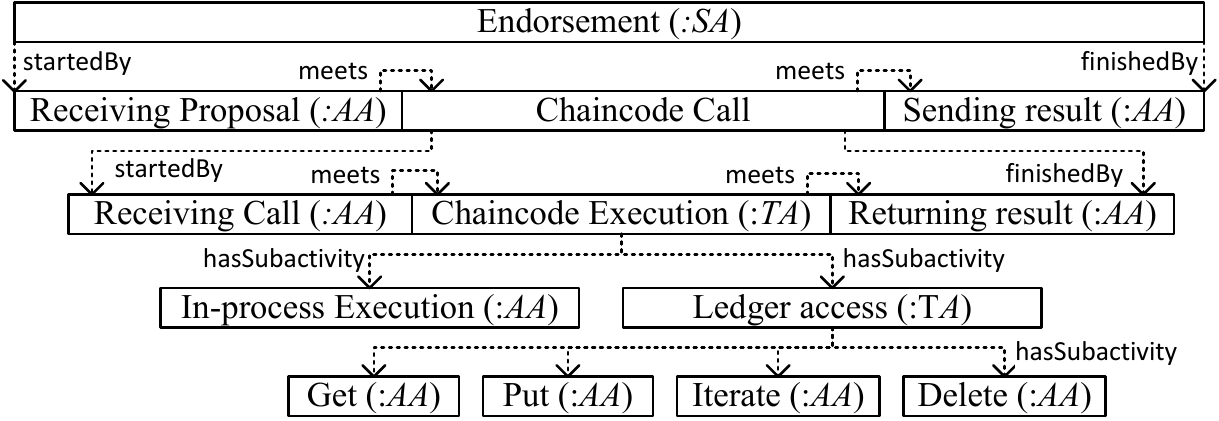}
\caption{Steps of the endorsement activity} \label{fig:hlf-endorsement}
\end{figure}

The endorsement activity (Fig.~\ref{fig:hlf-endorsement}) consists of the peer receiving the proposal, calling the required chaincode, then returning the result to the client. On the platform level, the \textit{Chaincode Call} activity type is not specified to enable refinement by different use cases, detailed in the next section.

\begin{figure}
\centering
\includegraphics[width=\textwidth]{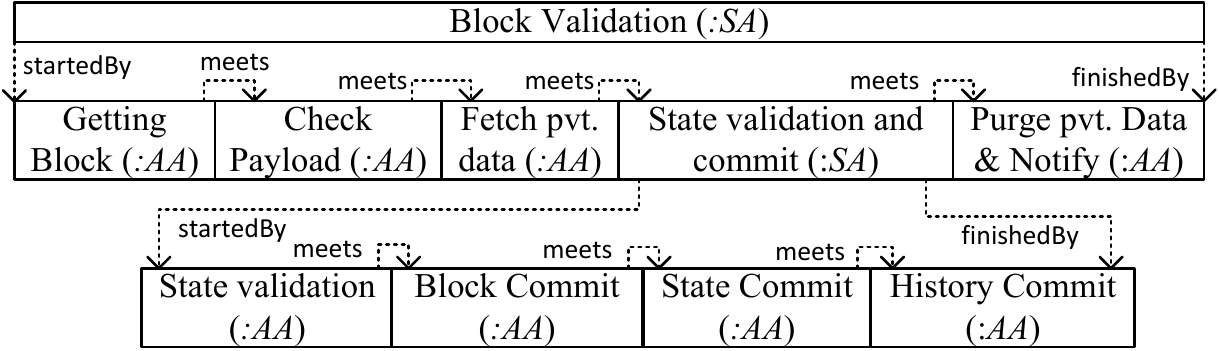}
\caption{Steps of the validation and commit activity} \label{fig:hlf-validation}
\end{figure}

The block validation and commit process of peers is modelled by a hierarchy of activity sequences (Fig.~\ref{fig:hlf-validation}). The validation begins by the ordering service delivering the new block to the peer (\textit{Getting Block}). Then the peer checks the block payload and fetches any private data (a privacy feature of HLF) required for further validation (\textit{Check Payload} and \textit{Fetch pvt. data} activities).

The \textit{State validation and commit} step is refined into further subactivities. First, the state modifications of transactions are validated (\textit{State validation}). Then the raw block content is committed to the blockchain storage (\textit{Block Commit}). Next the state modification of valid transactions are committed to the world state database (\textit{State Commit}). Finally, the history database is updated with the data accesses of committed transactions (\textit{History Commit}).

Finally, the peer purges stale private data and sends a notification about the block commit to subscribed clients. Once a client receives a notification about a block/transaction, the transaction life-cycle is considered complete.

\subsection{TPC-C and HLF bridge ontology}

The case study contains a final ontology that maps/bridges the TPC-C and HLF concepts, achieving the ''TPC-C on HLF'' model. The mapping plays the role of the platform-workload bridge ontology in Fig.~\ref{fig:general-use-case-structure}.

On one hand, the TPC-C case study chaincode was instrumented to measure the raw execution time of the chaincode. This allows the observation of peer-chaincode communication activities and differentiate between in-process execution and ledger access times (lower part of Fig.~\ref{fig:hlf-endorsement}). The exact control flow of the chaincode is not modelled, alternating activities are used instead to focus only on the duration of subactivities, and not on their order.

On the other hand, the bridge also refines the \textit{Execute TX} class of the TPC-C ontology. Due to the \textit{Read-only TX} and \textit{Read-write TX} class hierarchy, the following equivalence axioms are enough to specify that the workload is executed on HLF: i) \textit{Read-write TX} $\equiv$ \textit{Transaction Processing} and ii) \textit{Read-only TX} $\equiv$ \textit{Query Processing} (which is a simplified version of transaction processing, containing only the endorsement activity hierarchy, without further ordering or validation).
\section{Systematic measurement data analysis}
\label{sec:systematic-analysis}

At this point, the workload and platform ontologies are combined, and the measured activities are ''flagged'' with the appropriate \textit{BM/DM/EM} observability classes. Inputting the model to an OWL-DL reasoner will propagate the measured activity aspects throughout the rest of the model by flagging activities with different inference rule classes. The added classes denote how the beginning, duration and end of a flagged activity can be calculated based on other activity observations. 

The added rule classes define \textit{relations} between the temporal data of different activities. The following subsections provide examples for how such relations can be exploited to:
\begin{enumerate}
    \item correlate and validate the distributed measurement data;
    \item derive further, directly not measured (i.e., indirect) temporal data;
    \item validate the conformance of measurement data to the activity model;
    \item and systematically guide the bottleneck analysis tasks.
\end{enumerate}

\subsection{Correlate and validate measurement data}

Online services today exhibit a shift towards micro-service architectures to facilitate different DevOps tasks (e.g., rapid continuous delivery and deployment) and increase certain extra-functional properties of systems (e.g., availability, maintainability, fault tolerance, scalability). Accordingly, an end user request will traverse many services and corresponding components during processing. The same phenomenon is inherently present in distributed, peer-to-peer systems, such as HLF. 

In most cases a unique correlation/trace identifier is associated with each request to facilitate its tracing across component boundaries. HLF, for example, associates a unique transaction identifier (TX ID) with each client request, calculated from the client's identity and the time the transaction was constructed. When network components provide logs about certain transaction steps, they also log the corresponding TX ID along with the trace data.

A prerequisite of reconstructing a detailed activity timeline of transactions is the collection and correlation of such distributed traces. Novel observability frameworks, such as OpenTelemetry,\urlfootnote{https://opentelemetry.io/} may provide means to collect traces across component boundaries. For example, services utilizing OpenTelemetry can also send the collected traces (as metadata) along with the requests to other system components. Even though such approaches can ''centralize'' trace collection to a certain level, it is a rather intrusive instrumentation choice, hindering adoption by existing systems (such as HLF\urlfootnote{https://hyperledger.github.io/fabric-rfcs/text/0000-opentelemetry-tracing.html}). 

Instead, many systems opt to provide request trace data utilizing their already existing logging capabilities. In this case, distributed transaction traces must be collected and correlated using a separate monitoring stack, which presents its own challenges (but at least it is separate from the core system functionality). Having a detailed activity model for distributed transaction processing (such as the HLF consensus process) can facilitate the correlation and availability check of traces.

\begin{figure}
\centering
\includegraphics[width=0.9\textwidth]{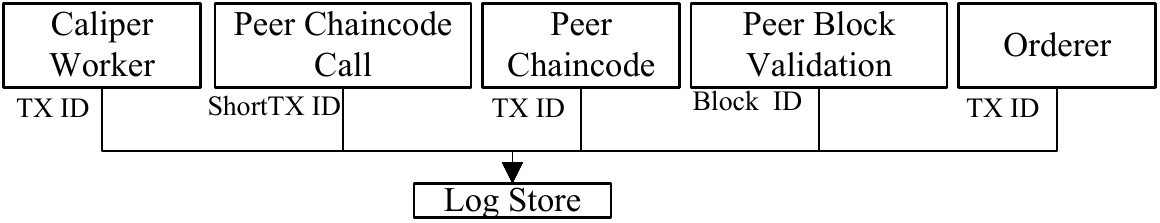}
\caption{Different trace sources of a transaction} 
\label{fig:hlf-trace}
\end{figure}

A HLF network setup usually contains the following trace sources (Fig.~\ref{fig:hlf-trace}):
\begin{itemize}
    \item optional end-to-end traces logged by the client (Caliper, in this case), with an associated TX ID;
    \item optional traces logged by chaincodes (one for each executing peer), with an associated TX ID;
    \item chaincode call traces logged by the peer nodes (one for each executing peer), with an associated \textit{shortened} TX ID (first $8$ characters only);
    \item block validation and commit traces logged by the peer nodes (one for each peer), with an associated block ID;
    \item and block creation traces logged by the leader orderer node, with an associated block ID.
\end{itemize}

The activity model of the case study defined the measured temporal data of activities, associated with the service types logging them. Accordingly, the prerequisite trace correlation step simply followed the structure of the model to check whether all supposedly measured data are available from all sources. 

The check revealed two anomalies:
\begin{enumerate}
    \item there were $4334$ transactions with missing traces;
    \item there were $2$ transactions with more traces than required.
\end{enumerate}

Case $1$ had an interesting symmetry in it: there were $2167$ transactions where \textit{all} Caliper-side traces were missing; and there were $2167$ transactions where \textit{all} other (non-Caliper) traces were missing. This lead to the hypothesis that one half is actually corresponding to the other half. 

Since the non-Caliper traces constituted a complete data set on their own, the focus of investigation was Caliper's Fabric integration. Further transaction metadata analysis revealed that all ''mismatched'' traces were HLF queries. Finally, the investigation revealed a \textit{bug in Caliper's query submitting logic}.\urlfootnote{https://github.com/hyperledger/caliper/issues/1187} Caliper created a TX ID for the request, but did not pass it along to the HLF SDK, which in turn created a new (and different) TX ID, unknown to Caliper. This resulted in client-side traces having a different TX ID than HLF-side traces.

Case $2$ was a similarly peculiar anomaly. Two transactions had chaincode call traces from peers that did not even execute those transactions. Closer inspection revealed that the shortened TX IDs contained a duplicate, i.e., two different TX IDs had the same shortened ($8$ characters) versions. Accordingly, the pairing of traces was not unique, two transaction got each others chaincode call traces. 

Luckily, the correct traces could be restored without data loss through temporal correlation: the ''conflicting'' transactions were executed well apart in time. However, if all peers would have executed those transactions, then the short TX ID conflict would have gone unnoticed until later in the analysis workflow (Sec.~\ref{subsec:data-validation}). The anomaly showed that reducing the information carried by trace correlation IDs is highly discouraged.

\subsection{Deriving indirect measurement data}
\label{subsec:deriving-data}

The final activity model of the HLF consensus refines a transaction into $28$ hierarchical steps even if only a single peer endorses and validates transactions. In general, the number of activities corresponding to a transaction is $5+13*E+10*V$, where $E \in \mathbb{N}^+$ is the number of endorsing peers for a transaction, and $V \in \mathbb{N}^+$ is the total number of peers in the network (since every peer validates transactions). 

Moreover, each activity has three associated temporal data: its beginning time, duration, and end time. Accordingly, the volume of temporal can quickly increase with the network size and the number of analysed transactions. For the sake of readability, let us assume that only a single peer endorses and validates transactions, resulting in $84$ potentially observable temporal data for the $28$ activities of \textit{each} transaction. 

\begin{figure}
\centering
\includegraphics[width=0.6\textwidth]{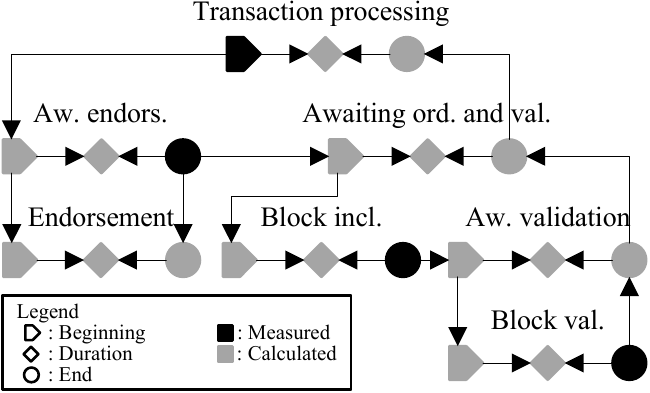}
\caption{Measurement propagation for high-level HLF activities} \label{fig:hlf-high-level-obs}
\end{figure}

\begin{figure}
\centering
\includegraphics[width=0.8\textwidth]{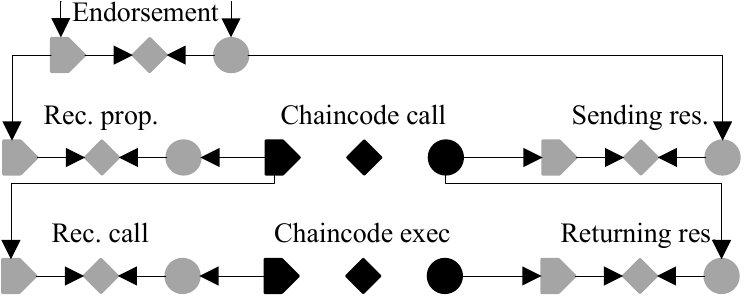}
\caption{Measurement propagation for the endorsement-related activities} \label{fig:hlf-end-obs}
\end{figure}

\begin{figure}
\centering
\includegraphics[width=\textwidth]{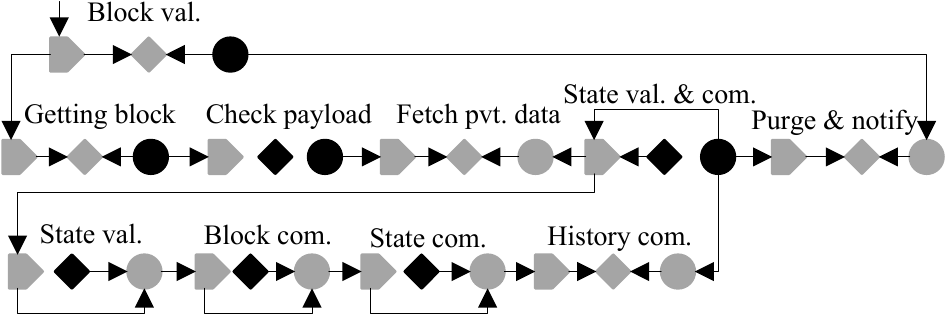}
\caption{Measurement propagation for the correct validation-related activities} \label{fig:hlf-val-obs-good}
\end{figure}

Figs.~\ref{fig:hlf-high-level-obs}--\ref{fig:hlf-val-obs-good} depict each activity and their corresponding temporal data (beginning, duration, and end). Black-filled shapes mark the directly measured data points. Using a component-off-the-shelf (COTS) HLF as SUT and Caliper as workload generator, there are $18$ directly measured data points:
\begin{itemize}
    \item Caliper marks: the beginning of a transaction; the end time when all endorsements arrive; and the end time when a notification is received about a committed block/transaction.
    
    \item Orderer nodes mark the end time when a new block is created.
    
    \item Peer nodes mark: the beginning, duration and end of a chaincode call; the end time when a block is received from an orderer; the end time and duration for checking the payload of a new block; and the end time and duration (including durations of some substeps) for validating and committing a block.
    
    \item The TPC-C chaincode implementation marks the start time, duration, and end time of the actual chaincode program execution.
\end{itemize}

The arrows in Figs.~\ref{fig:hlf-high-level-obs}--\ref{fig:hlf-val-obs-good} symbolize the direction of measurement data propagation, i.e., $A \longrightarrow B$ means that data $B$ is calculated from data $A$ (and possibly from other data in cases like $A \longrightarrow B \longleftarrow C$). The arrows essentially represent inference rules in the model, e.g., stating that the start time of an activity can be calculated from its end time and duration (like in the case of the \textit{State validation and commit} activity).

As shown in the figures, the directly measured temporal aspects are sufficient to completely observe the entire activity hierarchy through measurement propagation. If that were not the case, then the ''broken/missing'' data propagation paths would identify the places where the SUT needs additional sensor instrumentation to allow for more detailed observability.   

Note that Figs.~\ref{fig:hlf-high-level-obs}--\ref{fig:hlf-val-obs-good} are just a single, simplified view of a more complex data flow network determined by the applicable inference rules. The rigorous formal analysis of such data flow networks (in the context of temporal data) is subject to future work.

Moreover, the example assumes a single-peer HLF network. If the network consists of more than one peer, then the single-peer assumption is achieved by reducing the replicated endorsement and validation activities to a single instance by disregarding the non-bottleneck instances:
\begin{enumerate}
    \item Since transaction endorsements have a $WaitForAll$ synchronization semantic, keep only the longest running (i.e., the slowest) $Endrosement$ activity and its subactivities.
    
    \item Since block validations have a $WaitForAny$ synchronization semantic, keep only the shortest running (i.e., the fastest) $BlockValidation$ activity and its subactivities.
\end{enumerate}

At this point, a data analyst can use the formal data flow network to systematically derive new temporal data about the SUT's activities. However, an additional validation step is still needed to ensure not only the cleanness and richness, but also the correctness of the measurement data (or the model).

\subsection{Model and measurement data validation}
\label{subsec:data-validation}

Validating the measurement data is an important step to ensure the correctness of data analysis findings and insights. The proposed model-guided approach necessitates the following validation steps before proceeding to the performance analysis tasks:
\begin{enumerate}
    \item checking the conformance of measurement data to the activity model;
    \item and checking the consistency of the measurement data itself.
\end{enumerate}

\subsubsection{Detecting modelling errors}

The following scenario demonstrates how model conformance checks can reveal activity modelling errors. Such errors can be common if the model is reverse-engineered by others than the platform developers (like in this case study).

For example, HLF peers log the \textit{State validation and commit} activity details using the following message format: \texttt{[mychannel] Committed block ... in 26ms (state\_validation=3ms block\_and\_pvtdata\_commit=16ms state\_commit=3ms)}. 

\begin{figure}
\centering
\includegraphics[width=\textwidth]{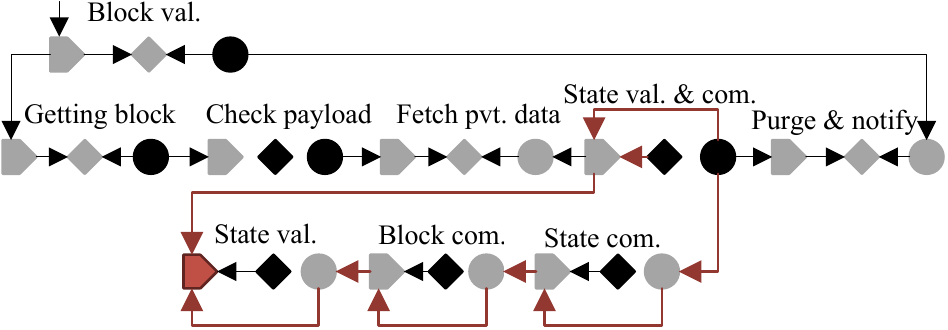}
\caption{Measurement propagation for the initial validation-related activities} \label{fig:hlf-val-obs-bad}
\end{figure}

Accordingly, a previous version of the consensus model refined the \textit{State validation and commit} activity as having only three subactivities (state validation, block commit, and state commit, as indicated by the log format). Fig.~\ref{fig:hlf-val-obs-bad} shows the temporal data propagation for the initial version.

Note how the (directly unobserved) beginning time of the \textit{State validation} subactivity can be calculated in two different ways (highlighted arrows in Fig.~\ref{fig:hlf-val-obs-bad}): i) based on sibling activity data; ii) and/or directly from parent activity data. There should not be any difference between the two paths in the case of a correct model and instrumentation. Validating this assumption requires checking whether the beginning times of the \textit{State validation} subactivities coincide with the beginning times of the \textit{State validation and commit} parent activities for every transaction, as required by the $startedBySubactivity$ relation among the two activity classes. 

\begin{figure}
\centering
\includegraphics[width=\textwidth]{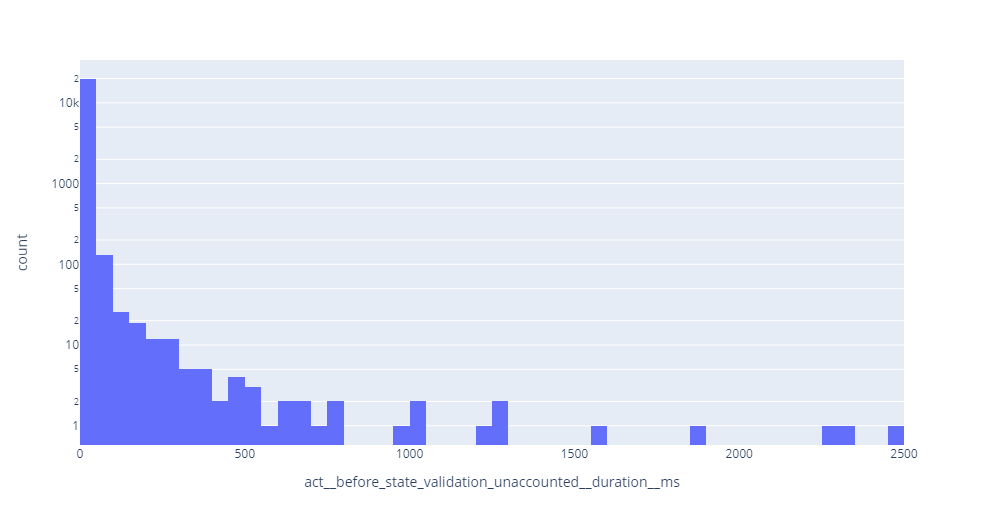}
\caption{Frequency distribution of missing validation time durations}
\label{fig:missing-time-state-validation}
\end{figure}

However, performing the check revealed that the equality constraint was violated for \textit{every} transaction. The \textit{State validation} activities always started later than their parent activities, indicating the presence of a hidden subactivity. Moreover, the magnitude of the missing time was sometimes non-negligible (Fig.~\ref{fig:missing-time-state-validation}), i.e., it could not be considered a measurement noise, thus warranting further investigation. 

As it turns out, the format of the log message was misleading and not all relevant subactivities were listed in the message. The source code inspection of HLF revealed that there is an other non-negligible subactivity performed during \textit{State validation and commit}, namely committing the state modifications of a transactions to a history database. Accordingly, the final model of the HLF consensus was extended with the \textit{Commit history} subactivity (Fig.~\ref{fig:hlf-val-obs-good}).

\begin{figure}
\centering
\includegraphics[width=\textwidth]{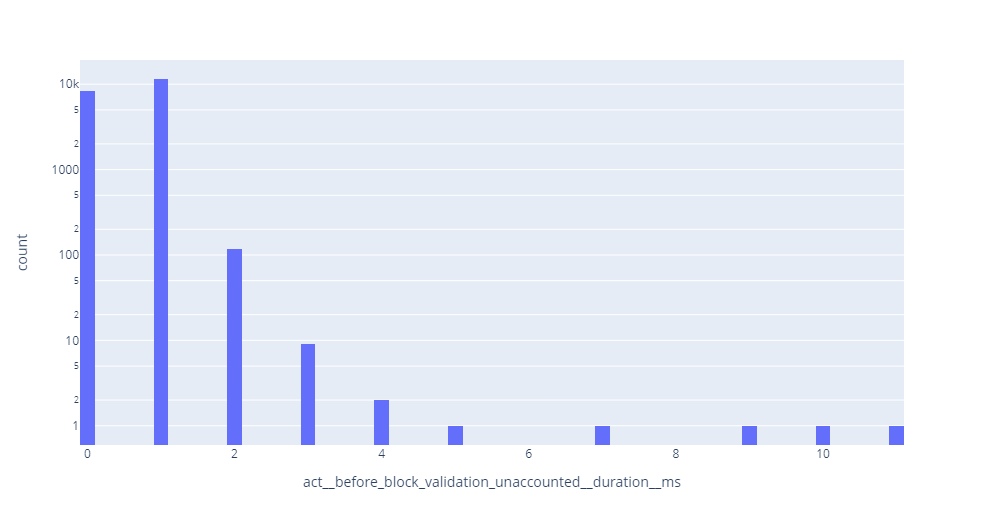}
\caption{Frequency distribution of missing payload check time durations}
\label{fig:missing-time-payload-check}
\end{figure}

Note that measurement noises are a common occurrence in complex, especially high-througput or overloaded systems. The measurement data conformance check also revealed some inconsistencies around the \textit{Check payload} activity. Calculating the beginning time of the activity from its own end time and duration yielded a different result than propagating the end time of its immediate predecessor \textit{Getting block} activity. Even though the propagation path is short and simple, it still violated the modeled activity relationship. However, the magnitude of ''missing'' times (Fig.~\ref{fig:missing-time-payload-check}) is negligible. 

One probable explanation could be that the missing time is a side-effect of the logging mechanism: the measured duration was calculated based on times $start_{calc}$ and $end_{calc}$, while the logging library marked the log message with an $end_{log} > end_{calc}$ timestamp, and $end_{log}$ was taken as the measured end time by the log processing pipeline. An other explanation could be that negligible activities were performed between the two modeled activities that can be safely ignored during performance analysis.

\subsubsection{Detecting measurement errors}

The systematic data propagation can also aid the detection of measurement (or measurement setup) errors. The missing subactivity issue manifested itself as ''missing time'' in the transaction timeline. The other important symptom of inconsistent measurement data is negative durations. 

The analysis showed negative \textit{Receiving proposal} activity durations upon measurement data validation. The duration in question is a derived measurement. Its value is indirectly calculated as the difference between the beginning time of calling a chaincode (\textit{Chaincode call} activity) and the beginning time of creating a transaction (\textit{Transaction processing} activity), both data being direct measurements. A negative duration result would mean that the chaincode is called before the transaction is even constructed, which is a serious \textit{event causality violation}. 

\begin{figure}
\centering
\includegraphics[width=\textwidth]{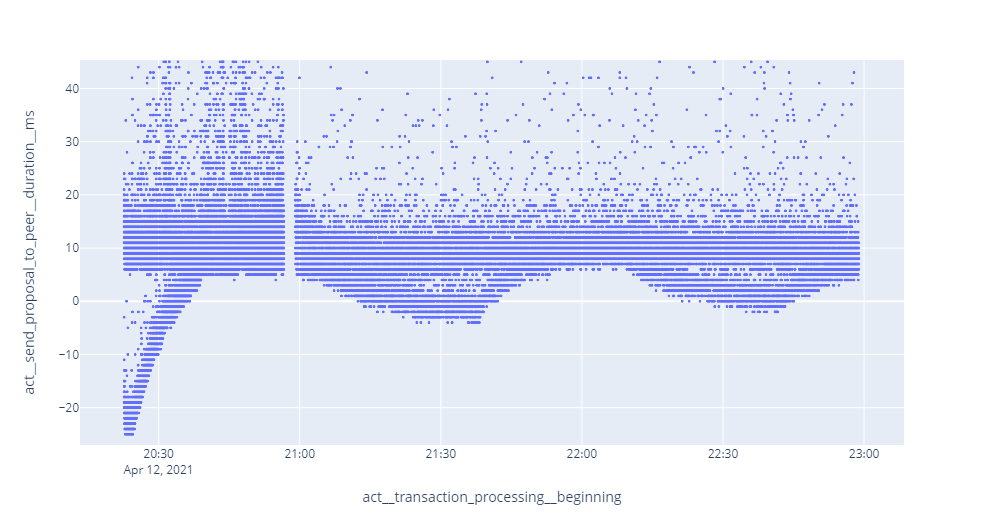}
\caption{Effect of misaligned system clocks over time}
\label{fig:clock-sync-issue}
\end{figure}

Note that the two direct measurements (the bases of the duration calculation) originate from two different (physical) components in the distributed network: the beginning of \textit{Transaction processing} is captured by Hyperledger Caliper (i.e., the HLF client), while the beginning of the {Chaincode call} is logged by the HLF peer nodes. Fig.~\ref{fig:clock-sync-issue} shows the \textit{Receiving Proposal} durations for each transaction over the time of the SUT measurement and reveals a curious trend: the anomalous durations smoothly oscillate around zero over time, i.e., negative durations are not that isolated and sporadic. Moreover, Fig.~\ref{fig:clock-sync-issue} depicts the activity data of each transaction after non-bottleneck endorsement activities have been eliminated, as outlined in Sec.~\ref{subsec:deriving-data}. Correspondingly, different \textit{Receiving Proposal} activity durations may originate from different peer nodes of the network. 

Combining the observations with the outlined assumptions results in the following working hypothesis: the system clock of a peer node periodically drifts out of sync from the other components. Measurement setup investigations later revealed that network nodes used a default, light-weigh time synchronization service instead of a more sophisticated one that provides higher precision. 

Measurement errors of such a low magnitude was deemed negligible in the previous section (Fig.~\ref{fig:missing-time-payload-check}). However, in this case, the presence of event causality violations shadows the usually insignificant magnitude of the actual measurement error. For example, process mining approaches can produce significantly different results in the presence of such causality violations. 

\begin{figure}
\centering
\includegraphics[width=\textwidth]{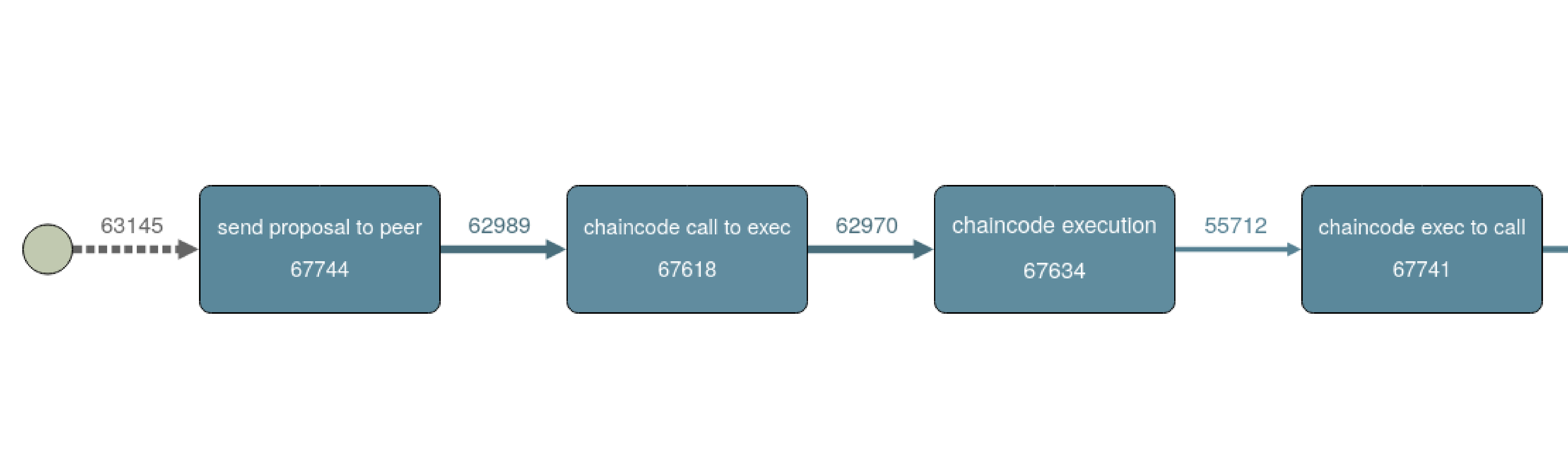}
\caption{Process mining result without causality violations}
\label{fig:hlf-process-good}
\end{figure}

\begin{figure}
\centering
\includegraphics[width=\textwidth]{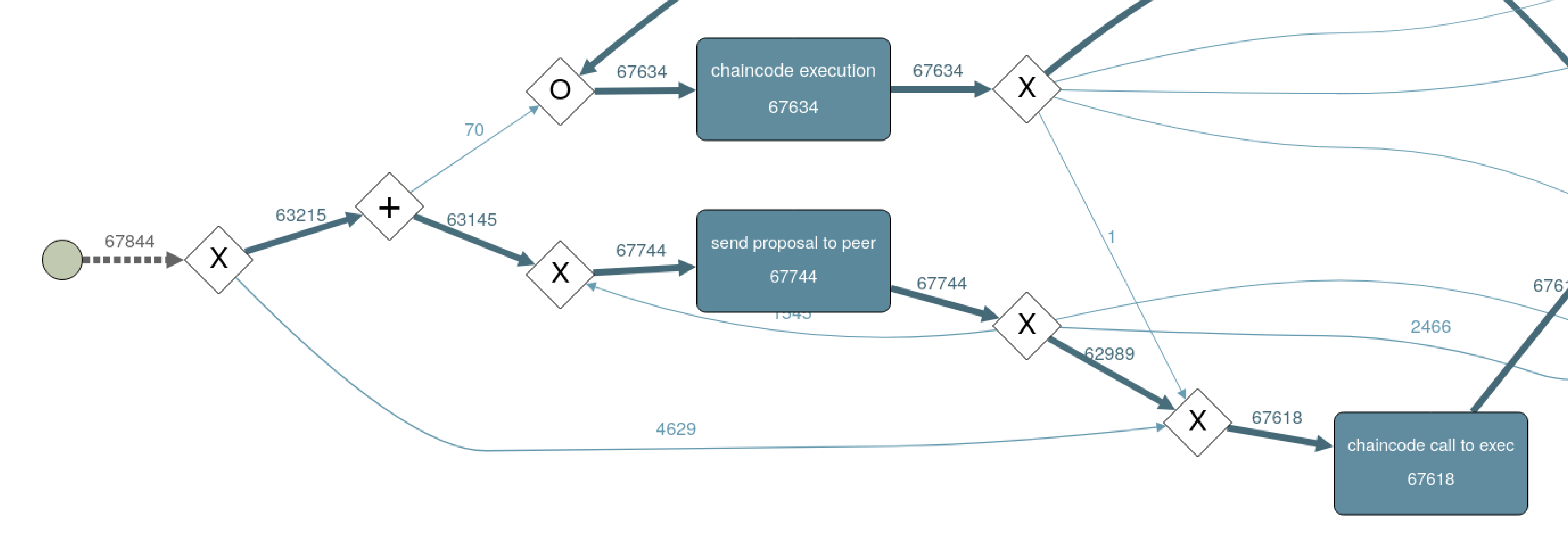}
\caption{Process mining result with causality violations}
\label{fig:hlf-process-wrong}
\end{figure}

Considering only the atomic activites of the HLF consensus model results in the low-level sequence of steps of the transaction life-cycle. Inputting the measurement data of such activities into a process mining algorithm should result in the process of Fig.\ref{fig:hlf-process-good}, assuming that the measurement data reflects the correct causality of events. However, the presence of causality violations in the input temporal data can lead to an incorrect process model (Fig.~\ref{fig:hlf-process-wrong}). Such models can hinder the correct understanding and insights of the SUT (that would be the goal of process mining) even for experienced HLF domain experts. 

Correspondingly, systematically cleaned and validated data is a must if the data analysis workflow incorporates formal approaches. The proposed approach and supporting ontology models enable rigorous (and possibly automated) measurement data validation before performing further performance analysis tasks.

\subsection{Guided bottleneck identification}
\label{subsec:bottleneck-id}

The primary goal and advantage of the proposed approach is that by the time the data analysts reach the actual performance analysis task, the available measurement data is validated, cleaned, and structured among semantically precise relations. The last section demonstrates how bottleneck identification and the root cause analysis of latency anomalies become intuitive and easy-to-perform tasks, given the proper input data.  

\begin{figure}
\centering
\includegraphics[width=\textwidth]{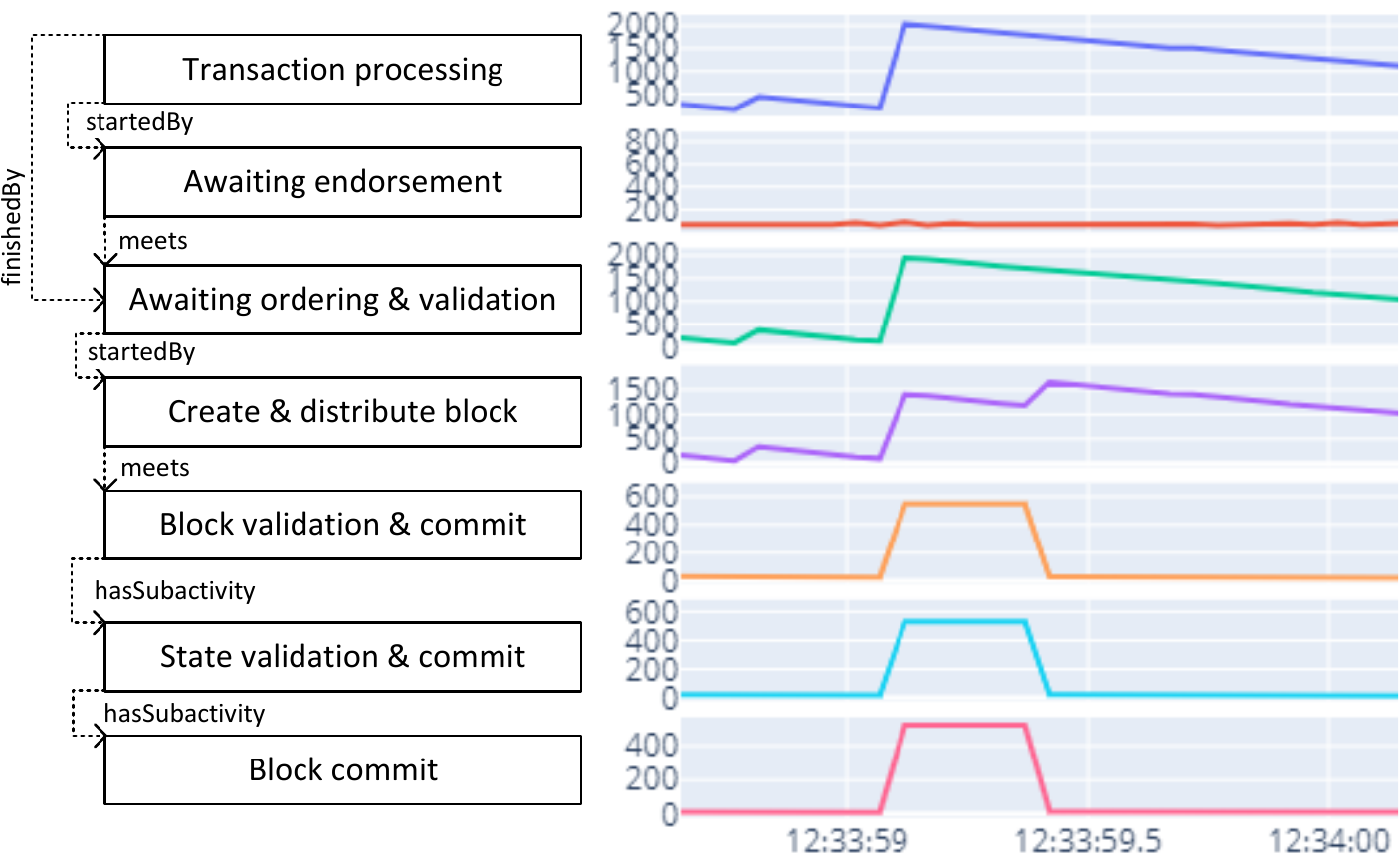}
\caption{A partial HLF activity time series hierarchy demonstrating a latency anomaly.} \label{fig:fabric-latencies}
\end{figure}

Let us assume that an end-to-end latency spike is detected on the client-side, classified as an anomaly (the exact anomaly detection methods are out of the scope of this paper). Fig.~\ref{fig:fabric-latencies} demonstrates how the hierarchical activity data aids the root cause analysis of the anomaly, uncovering bottlenecks contributing to the latency spike. 

The analysis employs a drill-down approach using the parent-subactivity hierarchy relations to gradually pinpoint significant latency contributors. At first, the latency of the high-level transaction processing subactivties are considered. Since endorsement times seem constant during the anomaly, the endorsement activity is dismissed as bottleneck and root cause. The ordering and validation subactivity, however, exhibits the same latency trend as the end-to-end anomaly. Correspondingly, it becomes the next activity of interest.

At this point, the subactivity latency trends show an interesting pattern. Neither the block creation, nor the block validation subactivites show the same trend as their anomalous parent activity. However, both indicate deviation from their previous baseline latency characteristics. Accordingly, the hierarchical exploratory process supports the identification of \textit{multivariate root causes}.

Block creation is a leaf activity element in the HLF consensus model, thus further root cause analysis along this path would require additional instrumentation or the detailed inspection of corresponding computing resource utilizations. The other prominent root cause path is the block validation activity. Further drill-down steps reveal that the atomic block commit activity caused the latency spike in this path. However, it must be noted that while the block commit anomaly is a short transient spike, the block creation latency needs more time to settle, hinting at some system statefullness and \textit{memory in the performance domain} (probably resulting from a queuing mechanism).

Nevertheless, the hierarchical and systematic approach allows the intuitive and quick identification of bottleneck activities of the SUT. Given the activities of interest, the next analysis steps include the correlation of bottleneck activity latencies with the corresponding component \textit{resource utilizations}, or with the \textit{characteristics of the workload}. Such correlations can answer the question whether the anomaly is caused by resource limits, or it is not really an anomaly, but a change in the presumed workload affected the exprected performance characteristics of the SUT. Such analysis, however, is outside the scope of this paper.

\section{Conclusion}
\label{sec:conclusion-and-future-work}

The increasing volume and dimensionality of performance measurement data necessitate the rigorous model-based support of data analysis tasks, such as bottleneck identification. While traditional DevOps approaches already benefit from MDE, performance data analysis lacks such support. 

The paper proposed an ontology-guided workflow (and presented the corresponding ODK) for modeling the composition of complex platform activities and their explicit observability. The ODK also supplies numerous inference rules to reason about the implicit observability of activities, creating a rich model serving as a strong formal basis for later performance analysis tasks. 

A representative case study demonstrated the advantages of the approach: a model-guided drill-down bottleneck identification process for a TPC-C benchmark workload executed on a HLF network. The current work aims at the integration of domain-specific knowledge in performance analysis into a core ontology, providing a strong formal foundation for measurement data analysis and performance monitoring of complex systems.

\bibliographystyle{actaplaindoi}
\bibliography{main}

\end{document}